\documentclass[%
 reprint,
 amsmath,amssymb,
 aps,
]{revtex4-2}
\usepackage{hyperref}
\usepackage{graphicx}
\usepackage{dcolumn}
\usepackage{bm}
\usepackage{comment}
\usepackage[dvipsnames]{xcolor}
\usepackage{siunitx}

\begin{document}

\preprint{APS/123-QED}


\title{Improved Receiver Noise Calibration for ADMX Axion Search: 4.54 to 5.41 $\si{\micro{eV}}$}
\author{M. Guzzetti}
\altaffiliation{Correspondence to mguzz28@uw.edu}
 \author{D. Zhang}
  \altaffiliation{Correspondence to dzhang95@uw.edu}
\author{C. Goodman}
\author{C. Hanretty}
\author{J. Sinnis}
 \author{L. J Rosenberg}
  \author{G. Rybka}
  \affiliation{University of Washington, Seattle, Washington 98195, USA}
   
\author{John Clarke}
\author{I. Siddiqi}
  \affiliation{University of California, Berkeley, California 94720, USA}
\author{A. S. Chou} 
 \author{M. Hollister} 
    \author{S. Knirck}
\author{A. Sonnenschein} 
  \affiliation{Fermi National Accelerator Laboratory, Batavia, Illinois 60510, USA}

\author{T. J. Caligiure}
\author{J. R. Gleason}
\author{A. T. Hipp}
\author{P. Sikivie}
\author{M. E. Solano}
\author{N. S. Sullivan}
\author{D. B. Tanner}
\affiliation{University of Florida, Gainesville, Florida 32611, USA}

\author{R. Khatiwada}
\affiliation{Illinois Institute of Technology, Chicago, Illinois 60616, USA}
\affiliation{Fermi National Accelerator Laboratory, Batavia, Illinois 60510, USA}

\author{G. Carosi}
\author{N. Du}%
\author{C. Cisneros}
\author{N. Robertson}
\author{N. Woollett}
\affiliation{Lawrence Livermore National Laboratory, Livermore, California 94550, USA}

\author{L. D. Duffy}
  \affiliation{Los Alamos National Laboratory, Los Alamos, New Mexico 87545, USA}

\author{C. Boutan}
\author{T. Braine}
\author{N. S. Oblath}
\author{M. S. Taubman}
\author{E. Lentz}
  \affiliation{Pacific Northwest National Laboratory, Richland, Washington 99354, USA}

\author{E. J. Daw}
\author{C. Mostyn}
  \author{M. G. Perry}
  \affiliation{University of Sheffield, Sheffield S10 2TN, UK}

\author{C. Bartram}
\author{T. A. Dyson}
\author{C. L. Kuo}
\author{S. Ruppert}
\author{M. O. Withers}
\author{A. K. Yi}
\affiliation{SLAC National Accelerator Laboratory, 2575 Sand Hill Road, Menlo Park, California 94025, USA}

\author{B. T. McAllister}
\affiliation{Centre for Astrophysics and Supercomputing,
Swinburne University of Technology}

\author{J. H. Buckley}
\author{C. Gaikwad}
\author{J. Hoffman}
\author{K. Murch}
\author{J. Russell}
  \affiliation{Washington University, St. Louis, Missouri 63130, USA}
  \author{M. Goryachev}
  \author{E. Hartman}
\author{A. Quiskamp}
\author{M. E. Tobar}
\affiliation{University of Western Australia, Perth, Western Australia 6009, Australia}

\collaboration{ADMX Collaboration}

\date{\today}

\begin{abstract}

Axions are a well-motivated candidate for dark matter. The preeminent method to search for axion dark matter is known as the axion haloscope, which makes use of the conversion of axions to photons in a large magnetic field. Due to the weak coupling of axions to photons however, the expected signal strength is exceptionally small. To increase signal strength, many haloscopes make use of resonant enhancement and high gain amplifiers, while also taking measures to keep receiver noise as low as possible such as the use of dilution refrigerators and ultra low-noise electronics. In this paper we derive the theoretical noise model based on the sources of noise found within a typical axion haloscope receiver chain, using the Axion Dark Matter eXperiment (ADMX) as a case study. We present examples of different noise calibration measurements at 1280~MHz taken during ADMX's most recent data-taking run. These new results shed light on a previously unidentified interaction between the cavity and JPA, as well as provide a better understanding of the systematic uncertainty on the system noise temperature used in the axion search analysis for this data-taking run. Finally, the consistency between the measurements and the detailed model provide suggestions for future improvements within ADMX and other axion haloscopes to reach a lower noise temperature. 

\end{abstract}

\maketitle

\section{Introduction}

Astrophysical observations indicate that 85\% of the matter content of the universe is in the form of non-baryonic dark matter \cite{Planck_2018}. Though there are numerous pieces of evidence pointing to the existence of dark matter, its fundamental nature remains one of the biggest mysteries to explore in modern physics. One particularly compelling dark matter candidate is the ``invisible'' axion, which was first proposed as a solution to the strong-CP problem~\cite{StrongCP_Problem,PRESKILL_1983_vacuum_realigment}, a puzzle in particle physics related to the unexpectedly small observed neutron electric dipole moment. Interestingly, axions produced in large quantities via the vacuum misalignment mechanism or the decay of the topological defects in the early universe could comprise all the local dark matter density~\cite{PRESKILL_1983_vacuum_realigment,ABBOTT_1983_vacuum_realignment,DINE_1983_vacuum_realignment}. 

The axion haloscope was proposed by Pierre Sikivie~\cite{Sikivie1983} to search for invisible axion dark matter in the 1980s and remains one of the most sensitive experimental designs  in the microwave regime. A microwave cavity is immersed in a strong magnetic field~($B$), and axion dark matter is converted into photons via the inverse Primakoff effect~\cite{primakoff}.
The energy of the photon carries the total energy of the axion, which is roughly equal to its rest mass due to the axion's small expected kinetic energy. The microwave cavity provides signal strength enhancement while the photon frequency is close to the cavity resonant frequency. The fundamental transverse magnetic~(TM) mode, TM$_{010}$, of the cavity is often used because it maximizes the signal strength by having the largest overlap between its electric field and the external $B$ field.

The sensitivity of the haloscope is limited by thermal noise introduced by the components closest to the cavity including the cavity itself and the electronics in the receiver chain. In the event of discovery, noise calibration directly affects the reported uncertainty on the axion-to-photon coupling $\times$ local dark matter density. With the use of quantum amplifiers operating at milliKelvin temperatures and providing larger than 15~dB gain~\cite{RobertMSAthesis,msa,jpa1,jpa2,twpa}, haloscopes are able to approach the standard quantum limit~(SQL). The electronic noise is dominated by the first stage amplifier, which is why haloscopes typically situate an ultra-low-noise quantum amplifier at the beginning of the receiver chain~\cite{admx2021,capp_2024,haystac2023}.

Some novel techniques to reach a noise level lower than the SQL such as photon counting, quantum squeezing or the state-swapping interaction~\cite{photonCounting,squeezing,stateSwapping} are beyond the scope of this paper. 



The system noise $T_{\rm sys}$ calibration of a haloscope can be done by the Y-factor measurement with a variable temperature stage (VTS), which shares the same receiver chain with the cavity after a cryo-switch (see Fig.~\ref{fig:rf_layout})~\cite{jpaY-factor2021}. The relationship between output power and a VTS~\cite{ParametricAmpCalibration} created by a resistor-heated noise source is used to infer the system noise temperature. However, the gain and noise contribution of the quantum amplifiers are sensitive to environmental factors including temperature, mechanical vibration and detune frequency. Plus, the Y-factor measurement often takes a couple hours of downtime due to the time it takes for the VTS to reach an equilibrium temperature while heating and cooling. Additionally, due to the narrowband response of some quantum amplifiers (such as the Josephson Parametric Amplifier (JPA) that is used by ADMX), the calibration can be done at only one frequency at a time. Therefore, it's challenging to use the direct Y-factor measurement of the system with active quantum amplifiers to provide a timely update of $T_{\rm sys}$ for haloscopes.

The signal-to-noise-ratio improvement~(SNRI) method (see details in Sec.~\ref{sec:admxModel}) is able to monitor $T_{\rm sys}$ \textit{in-situ} in combination with the Y-factor measurement results from when the quantum amplifier is inactive,  $T_{\rm sys,off}$~\cite{run1bdetail}. The noise contributed by the receiver chain after the quantum amplifier is more stable with the regular running condition changes including mechanical vibrations from the cryo-liquid fills and temperature oscillations in the insert space cooled to milliKelvin temperatures. Aside from $T_{\rm sys,off}$ which requires a full Y-factor calibration, the related quantities in SNRI can be measured within a minute for every data taking cycle, reflecting the difference between  $T_{\rm sys}$ and $T_{\rm sys,off}$ by measuring the absolute output power differences and the quantum amplifier gain. 

In addition to this traditional Y-factor technique, there are other, novel, noise calibration methods such as a Y-factor measurement using a shot noise tunnel junction (SNTJ)~\cite{ParametricAmpCalibration}, or a switchless tone-injection based Y-factor measurement~\cite{Braggio_2022}. While both of these methods greatly improve calibration speed when compared with a VTS Y-factor measurement, ADMX has not implemented them yet for a few reasons. Namely, SNTJs are not easily accessible or easily fabricated compared to a simple VTS, and the switchless method requires that the attenuation of the RF lines used for calibration and the cavity itself not be too large that they drown out the injected signals. Additionally, in the case of ADMX, our most recent data-taking run lasted almost a full year, so in the broader picture a noise calibration technique that takes a few hours in total does not significantly impact our downtime.

In this work, we first introduce the common components in the cryo-space contributing to the system noises. Secondly, we establish a detailed noise model for the Axion Dark Matter eXperiment~(ADMX). Thirdly, we demonstrate the noise calibration techniques 
currently employed in ADMX that can be easily accommodated to other axion haloscopes.  Specifically, we compare the direct Y-factor measurement of an active JPA using a VTS with the SNRI method at different frequencies. We also evaluate the Y-factor noise calibration results using the cavity as our noise source. Finally, we examine the noise calibration results as a whole, and discuss the differences between the various methods as well as the appropriate use of each result.

\section{Sources of Noise}

We are interested in our ability to discriminate a signal from thermal noise after it has passed through a series of radio-frequency~(RF) components.  These components may amplify or attenuate the signal while adding additional thermal noise.  
The signal-to-noise ratio (SNR) is a fundamental measure of the sensitivity, which is 
the ratio of the signal power $P_\mathrm{sig}$ to the noise power $P_\mathrm{sys}$ over some bandwidth $b$ at the output of our RF system, so we define the noise temperature $T_\mathrm{sys}$ of a device to be scaled relative to the input signal reference plane as
\begin{eqnarray}\label{eqn:snr}
    {\rm SNR} = \frac{P_\mathrm{sig,out}}{\sigma_\mathrm{sys,out}} = \frac{P_{\rm sig}}{k_{B}T_{\rm sys}}\cdot\sqrt{\frac{t}{b}},
\end{eqnarray}
where the radiometer equation~\cite{dicke} is implied, $k_{B}$ is the Boltzmann's constant, $t$ is the integration time of the digitization and $P_{\rm sig}$ refers to the signal power entering the detecting system which depends on the coupling of the antenna. For clarity, ``reference plane'' here defines the location in the RF chain where the quantities of interest ($T_\mathrm{sys}$ and the resulting SNR, in this case) are being defined from. That is, the system noise temperature and corresponding SNR are defined as above from the specified reference plane to the output of the receiver.



\subsection{Blackbody Noise}\label{sec:blackbody}
The noise temperature of a blackbody is a function of its physical temperature $T_{\rm phys}$ and the frequency $f$:
\begin{equation}
T_{\mathrm{noise}}(f, T_\mathrm{phys})= \frac{h f}{2k_{B}} \coth \left( \frac{hf}{2k_{B}T_\mathrm{phys}} \right),
\end{equation}
where $h$ is the Planck constant~\cite{Callen_Wellen_1951}.
The noise power is related to $T_{\mathrm{noise}}$ as 
\begin{equation}
P_\mathrm{noise} =  k_{B}b T_{\mathrm{noise}}.
\end{equation}
 While $T_{\rm phys}\gg hf/2k_{B}$, $T_{\mathrm{noise}}$ is approximately equal to $T_{\rm phys}$, corresponding to a thermally limited system. However, when $T_{\rm phys}\ll hf/2k_{B}$, $T_{\mathrm{noise}}={h f}/{2k_{B}}$ which is equal to the SQL.

\subsection{Passive Attenuator}\label{sec:att}
By Kirchhoff's law, the emissivity and absorptivity of a passive object in thermal equilibrium must be the same~\cite{Kirchoff}.
Consider a blackbody at a noise temperature $T_A$ and attenuator at noise temperature $T_B$ which transmits power fraction of $\alpha$ and absorbs a power fraction of ($1-\alpha$).  The attenuator will also radiate as a blackbody with emissivity ($1-\alpha$) \cite{ParametricAmpCalibration}. The noise power and temperature as measured downstream will be
\begin{eqnarray}
P_\mathrm{noise}&=&k_{B}b\left(\alpha T_{A}+(1-\alpha)T_{B} \right),
\end{eqnarray}
and 
\begin{eqnarray}
T_{\mathrm{noise}}&=&T_{A}+\frac{1-\alpha}{\alpha}  T_{B}
\end{eqnarray}
with the reference plane at $A$.

\subsection{Active Component}
Consider a blackbody with a noise temperature $T_A$ amplified by an amplifier with gain $G$ and a noise temperature $T_B$, followed by downstream components that introduce additional noise $T_C$.  The noise temperature from the downstream components will be suppressed by the gain of the amplifier \cite{FRIIS_noisefigures}, leading to 
\begin{eqnarray}
P_\mathrm{noise}&=&k_{B}b\textbf{(}G\left(T_A+T_B\right)+T_C\textbf{)},
\end{eqnarray}
and
\begin{eqnarray}
T_{\mathrm{noise}}&=&T_A+T_B+\frac{T_C}{G},
\end{eqnarray}
with the reference plane at $A$.

\subsection{Parametric Amplifier}

In detectors like ADMX, a quantum parametric amplifier is often used in the scattering mode of operation as the first-stage amplifier~\cite{clerk}.
Ideally, a four-way (three-way) mixing parametric amplifier pumped at the frequency $f_{P}$ ($f_{P}$/2) can reach zero excess noise other than the zero-point fluctuations at the signal $f_{S}$ and idler frequency $f_{I}$, where $f_{I} = 2 f_{P} -f_{S}$ ($f_{P} -f_{S}$)~\cite{ParametricAmpCalibration}. If the signal frequency has a noise temperature $T_S$ and the idler frequency has a noise temperature $T_I$, and the gain of the parametric amplifier is $G\gg 1$,  followed by downstream components that introduce additional noise $T_D$, the output noise temperature will be measured downstream as 
\begin{eqnarray}
P_\mathrm{noise}&=&k_{B}b\textbf{(}G\left(T_S+T_I\right)+T_D\textbf{)},
\end{eqnarray}
and
\begin{eqnarray}
T_{\mathrm{noise}}&=&T_S+T_I+\frac{T_D}{G}
\end{eqnarray}
with the reference plane at the input to the parametric amplifier.a 
In reality, the parametric amplifier in ADMX still adds noticeable extra noises $T_{\rm JPA}$, so the noise temperature becomes
\begin{eqnarray}
T_{\mathrm{noise}}&=&T_S+T_I+T_{\rm JPA}+\frac{T_D}{G}.
\end{eqnarray}


\subsection{Circulators}
A circulator is a three port device for which, over its operational band, signals incident on port 1 exit port 2, signals incident on port 2 exit port 3, and signals incident on port 3 exit port 1~\cite{PozarDavidM2012Me}. 
 We note the power transmissivity from port $i$ to port $j$ as $\alpha_{{\rm circ},ji}$. For an ideal circulator $\alpha_{{\rm circ},21}=1$, $\alpha_{{\rm circ},32}=1$ and $\alpha_{{\rm circ},13} = 1$ and all the other permutations have $\alpha_{{\rm circ},ji}=0$. Cryogenic microwave circulators have small but measurable losses, and can be treated as attenuators for the purposes of noise as described in Sec.~\ref{sec:att}.

\subsection{Microwave Cavity}

Axion haloscopes commonly use at least one microwave cavity, which has a resonant mode of interest with an unloaded quality factor $Q$ coupled to an antenna with coupling $\beta$.  For frequencies near a resonance of interest $f_0$,  power incident on the cavity is reflected with reflectivity
\begin{equation}
|\Gamma_{\rm cav}(f)|^2=1-\frac{4\beta}{\left(1+\beta\right)^2} \frac{1}{1+4Q_L^2\left(\frac{f-f_0}{f_0}\right)^2},
\end{equation}
where $Q_L = Q/(1+\beta)$ is the loaded quality factor~\cite{run1bdetail, PozarDavidM2012Me}.
For a critically coupled ($\beta=1$) cavity on resonance, $|\Gamma_{\rm cav}(f)|^2 = 0$ and the cavity appears as a blackbody radiating with  physical temperature of the cavity.  Otherwise, $|\Gamma_{\rm cav}(f)|^2 \neq 0$ and the noise temperature as seen from the antenna is a mixture of the cavity's noise temperature $T_\mathrm{cav}$ and power reflected off of the antenna $T_{\mathrm{incident}}$
\begin{equation}
T_{\mathrm{noise}}=T_{\mathrm{incident}} |\Gamma_{\rm cav}|^2+ T_{\mathrm{cav}} \left(1-|\Gamma_{\rm cav}|^2\right).
\end{equation}

\section{Example Haloscope Model}\label{Example Haloscope Model}
As an example, consider the recent ADMX RF system from Run 1B to Run 1D shown in Fig.~\ref{fig:rf_layout}.  The primary signal path is from the cavity, through two circulators (Circ1 and Circ2), amplified off of a four-way mixing JPA, through two more circulators (Circ2 and Circ3), amplified by a heterostructure field effect transistor amplifier (HFET), and then up to the warm receiver.  Thermal noise comes from the attenuator A, passes through Circ1 and reaches the antenna.  Depending on the frequency and coupling, some of this noise is reflected, and some is replaced by thermal noise from the cavity.  The noise then passes through the same path as the signal, where additional noise will be added by attenuation in the cables and circulators, by mixing with the idler frequency at the JPA, by the HFET amplifier and the post-amplifiers in the warm receiver.  
In practice, the idler frequency for the JPA is always many Q widths away from the cavity resonance so its noise can be treated as independent from the cavity temperature. More specifically, the measurements in Sec.~\ref{sec:measurements} always detune $f_p/2$ by 320~kHz higher to the cavity resonance ($f_S$) while the bandwidth of the cavity is 56~kHz. Also, the HFET gain is quite high (40 dB), so the downstream noise addition is insignificant compared to the HFET noise.

The system can be run with the JPA powered by a pump tone with stable gains up to 25~dB, or with the JPA inactive, in which case it behaves as an ideal reflector.  The switch S can be flipped so that the noise comes from the ``hot load'' (VTS used in ADMX) for calibration.  The cable length and composition between the hot load and the switch is designed to be the same as that between the cavity and the switch, so the attenuation can be treated as nearly the same.

\subsection{ADMX Noise Model}\label{sec:admxModel}
\begin{figure*}
    \includegraphics[width=\textwidth]{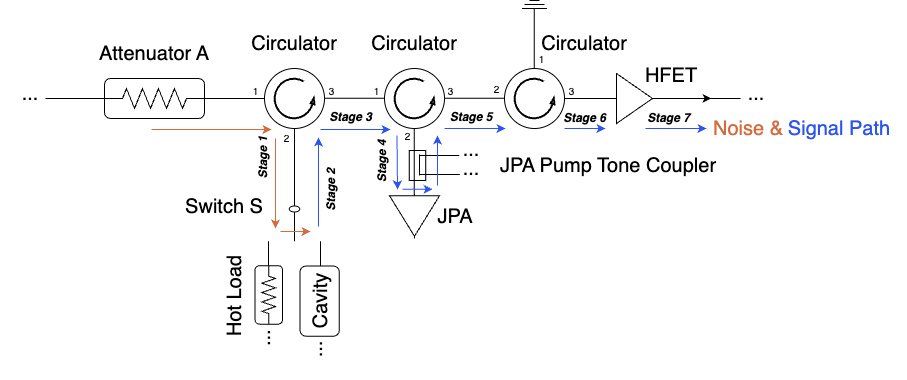}
    \caption{ADMX RF diagram in the cold space. Other than the HFET, all the components are connected to the milliKelvin stage directly. The blue arrows show the shared path for both noises and possible signal from the cavity. The brown arrows show the path of the attenuator A thermal noise to the cavity. Different stages (i.e. stage 1, stage 2, etc.) are labeled corresponding to the noise model in Sec.~\ref{sec:admxModel}.}
    \label{fig:rf_layout}
\end{figure*}

 We build the thermal model by assuming temperature gradients among all the critical cryo-components including the cavity, attenuator A, the hot load, etc.~(Fig.~\ref{fig:rf_layout}) even though, ideally, they should all be thermalized to the milliKelvin temperature stage, except the HFET. 
 The cavity, attenuator A, and the hot load are separately instrumented with temperature sensors which indicate corresponding blackbody noise temperatures $T_{\mathrm{cav}}$, $T_{\mathrm{A}}$, and $T_{\mathrm{HL}}$, respectively. 
 For the magnetic-field sensitive components  including the circulators, the switch and JPA that are mounted to a cold finger in a field-free region, a system we call the quantum amplifier package~\cite{magnetShield,run1a}, we start from assuming different physical temperatures at different circulators for generality, and later on, we simplify the model by using the fact that the components on the quantum amplifier package are thermalized to the same temperature $T_{\mathrm{circ}}$.
 The HFET noise temperature combined with any downstream receiver noise will be labelled $T_{\mathrm{HFET}}$. The gains of the JPA, the HFET and the equivalent downstream post amplifiers are noted as $G_{\rm JPA}$, $G_{\rm HFET}$ and $G_{\rm post}$, respectively.


The noise power with the JPA unpowered can be modeled by separating the cryo-space into different stages, where
\begin{eqnarray} \label{eqn:p_final}
P_{\rm stage1} &=& k_{B}bT_{ A} \alpha_{\rm{circ1},21}+ k_{B}bT_{\rm{circ1}} (1-\alpha_{\rm{circ1},21})\nonumber\\
P_{\rm stage2} &=& P_{\rm stage1}  |\Gamma_{\rm cav}|^{2} + k_{B}bT_{\rm cav}  (1-|\Gamma_{\rm cav}|^{2} )\nonumber\\
P_{\rm stage3} &=& P_{\rm stage2} \alpha_{\rm{circ1},32} + k_{B}bT_{\rm{circ1}} (1-\alpha_{\rm{circ1},32})\nonumber\\
P_{\rm stage4} &=& P_{\rm stage3} \alpha_{\rm{circ2},21} + k_{B}bT_{\rm{circ2}} (1-\alpha_{\rm{circ2},21})\nonumber\\
P_{\rm stage5} &=& P_{\rm stage4}  \alpha_{\rm{circ2},32} +k_{B}b T_{\rm{circ2}} (1-\alpha_{\rm{circ2},32})\nonumber\\
P_{\rm stage6} &=& P_{\rm stage5}  \alpha_{\rm{circ3},32} + k_{B}bT_{\rm{circ3}}(1-\alpha_{\rm{circ3},32})\nonumber\\
P_{\rm stage7} &=& G_{\rm HFET} (P_{\rm stage6} + k_{B}bT_{\rm HFET})\nonumber\\
P_{\rm noise,out} &=& G_{\rm post} P_{\rm stage7}.
\end{eqnarray}

While the JPA is on, we rewrite the relation between $P_{\rm stage4}$ and $P_{\rm stage5}$ as 
\begin{eqnarray} \label{eqn:p_jpa}
P_{\rm stage5} &=& \textbf{(}G_{\rm JPA} (P_{\rm stage4} +P_{{\rm JPA},S}) \nonumber \\
& &+(G_{\rm JPA}-1) (P_{\rm I} + P_{{\rm JPA},I})  \textbf{)}\alpha_{\rm{circ2},32} \nonumber \\
& &+ k_{B}b T_{\rm{circ2}} (1-\alpha_{\rm{circ2},32}), 
\end{eqnarray}
where $P_{{\rm JPA},S}$ and $P_{{\rm JPA},I}$ are extra noises due to the JPA at the signal and idler frequencies, respectively, due to the imperfect JPA amplifier. $P_{I}$ is the noise power at the idler frequency 
which can be traced up to $P_{\rm stage1}$. As the idler frequency is always off-resonance to the cavity, the reflection coefficient between stage 1 and stage 2 at the idler frequency is $1$.
More explicitly,
\begin{eqnarray}\label{equ:p_i}
    P_{I} &=& \textbf{(}P_{\rm stage1}\alpha_{{\rm circ1},32} + k_{B}bT_{\rm circ1 }(1-\alpha_{{\rm circ1},32}) \textbf{)} \alpha_{{\rm circ2},21} \nonumber \\ 
    & &+k_{B}bT_{\rm circ2 }(1-\alpha_{{\rm circ2},21}) .
\end{eqnarray}

If there is a signal power $P_{\rm sig}$ coming out of the cavity, $T_{\rm cav}(1-|\Gamma_{\rm cav}|^2)$ will be replaced with $\textbf{(}T_{\rm cav} (1-|\Gamma_{\rm cav}|^2) + P_{\rm sig}\textbf{)}$ in Eq.~\ref{eqn:p_final}. 
According to Eq.~\ref{eqn:snr}, we compare the $P_{\rm sig}$ to the system noise $T_{\rm sys}$ with the stage 2 as the reference plane because the signal comes into the receiver chain at the stage 2. 

If all the attenuation and amplifications at different stages are known, $T_{\rm sys}$ can be calculated as
\begin{eqnarray}\label{equ:tsys_snri}
   T_{\rm sys} = \frac{P_{\rm noise,out} }{ k_{B}b\alpha_{1}\alpha_{2} G_{\rm total}},
\end{eqnarray}
where $\alpha_1$ ($\alpha_{{\rm circ1},32} \alpha_{{\rm circ2},21}$) is the transmissivity from the cavity to the JPA, and $\alpha_2$ ($\alpha_{{\rm circ2},32}\alpha_{{\rm circ3},32}$) from the JPA to the HFET.
$G_{\rm total}= G_{\rm JPA} G_{\rm HFET} G_{\rm post} $ while the JPA pump is on and  $G_{\rm total}=  G_{\rm HFET} G_{\rm post} $ while off.

Even though it is difficult to have a direct accurate measurement of $G_{\rm post}$ during data taking, $G_{\rm post}$ can be cancelled out by comparing the noise powers with the JPA unpowered $T_{\rm sys,off}$ or powered $T_{\rm sys,on}$ because
\begin{eqnarray}\label{eqn:SNRI_tsys}
T_{\rm sys,on} = \frac{T_{\rm sys,off}}{\rm SNRI},
\end{eqnarray}
where SNRI refers to the signal-to-noise-ratio increase as
\begin{eqnarray}\label{eqn:SNRI}
{\rm SNRI}= \frac{G_{\rm JPA} P_{\rm noise,out,off}}{P_{\rm noise,out,on}}.
\end{eqnarray}
The noise power coming out of the system for either JPA on $P_{\rm noise,out,on}$ or off $P_{\rm noise,out,off}$ is measured timely, and so is $G_{\rm JPA}$.
To estimate $T_{\rm sys,on}$ with SNRI, ${T_{\rm sys,off}}$ has to be known first. Therefore, it's worthwhile to carefully trace both the JPA active and inactive model.

\subsection{Model Simplification}
We can simplify the noise model because all the circulators are thermalized to the same temperature $T_{\rm circ}$, and we preserve $\alpha_1$ and $\alpha_2$ introduced in Eq.~\ref{equ:tsys_snri}. 
In addition to Eq.~\ref{equ:tsys_snri}, $T_{\rm sys}$ can also be decomposed as follows.

When the JPA is off,
\begin{eqnarray}\label{eqn:tsysOff}
T_{\rm sys, off} &=& {T_{\rm stage1}} |\Gamma_{\rm cav}|^2  + T_{\rm cav} (1-|\Gamma_{\rm cav}|^2) \nonumber\\ 
& &+\frac{T_{\rm circ}(1-\alpha_1\alpha_2)}{\alpha_1\alpha_2}  +\frac{T_{\rm HFET}}{\alpha_1\alpha_2 },
\end{eqnarray}
where we convert the notations of the powers $P_{*}$ to the noise temperatures $T_{*} = P_{*}/k_{B}b$ for readability.

When JPA is on and $G_{\rm JPA} \gg 1$ is reached ($G_{\rm JPA} \approx G_{\rm JPA} - 1$), 
\begin{eqnarray}\label{eqn:tsysOn}
    T_{\rm sys, on} &=& {T_{\rm stage1}} |\Gamma_{\rm cav}|^2  + T_{\rm cav} (1-|\Gamma_{\rm cav}|^2) \nonumber\\ 
   & &+ {T_{\rm stage1}} + 2\frac{T_{\rm circ}(1-\alpha_1)}{\alpha_1}+
    \frac{T_{{\rm JPA}}}{\alpha_1} \nonumber \\
    & & + \frac{T_{\rm circ}(1-\alpha_2)}{\alpha_1\alpha_2 G_{\rm JPA}} +\frac{T_{\rm HFET}}{\alpha_1\alpha_2 G_{\rm JPA}}.
\end{eqnarray}
  Here we use $T_{\rm JPA}$ to denote the total extra noise introduced by the JPA which is equal to the sum at both the signal and idler frequencies $(T_{\rm JPA,S} + T_{\rm JPA,I})$ since the two noises are not separable.

\subsection{Hot Load Case}
To calibrate the noise temperature, a hot load with a 50~$\Omega$ terminator~(a typical reactance used for RF transmission lines) can be connected into the system with an RF switch as shown in Fig.~\ref{fig:rf_layout}.
When we switch to the hot load configuration from the cavity, $T_{\rm cav}$ is replaced with $T_{\rm HL}$, and $\Gamma_{\rm cav}$ with $\Gamma_{\rm HL}=0$ in Eqs.~\ref{eqn:tsysOff} and \ref{eqn:tsysOn}.

When the JPA is off, the $T_{\rm sys}$ becomes 
\begin{eqnarray}\label{eqn:hljpaoff}
\mathrm{T_{\rm sys,off,HL} }&=&T_{\rm HL}  \nonumber\\
& &+\frac{T_{\rm circ}(1-\alpha_1\alpha_2)}{\alpha_1\alpha_2}  +\frac{T_{\rm HFET}}{\alpha_1\alpha_2 }.
\end{eqnarray}

When the JPA is on,
\begin{eqnarray}\label{eqn:hljpaon}
    T_{\rm sys, on,HL} &=& 2 T_{\rm HL} \nonumber\\ 
   & &+ 2\frac{T_{\rm circ}(1-\alpha_1)}{\alpha_1}+
    \frac{T_{{\rm JPA}}}{\alpha_1} \nonumber \\
    & & + \frac{T_{\rm circ}(1-\alpha_2)}{\alpha_1\alpha_2 G_{\rm JPA}} +\frac{T_{\rm HFET}}{\alpha_1\alpha_2 G_{\rm JPA}}  .
\end{eqnarray}

\subsection{Cavity Cool-Down And Warm-Up Case}
When the system is thermalized to the same temperature from the A connected to Circ1 (stage 1) to the signal at the input of the HFET (stage 6), some terms in the $T_{\rm sys}$ model will cancel out and the equation will simplify. The receiver chain is thermalized in this way when the entire system is cooling down or warming up with respect to the same mixing chamber temperature ($T_{\rm mxc}$). These simplifications require that one is fully off-resonance since the cavity often takes more time to thermalize.

When the JPA is off,
\begin{eqnarray}\label{eqn:mxcjpaoff}
    T_{\rm sys,off,mxc} &=& \frac{T_{\rm mxc} +T_{\rm HFET} }{\alpha_1\alpha_2}.
\end{eqnarray}

When the JPA is on,
\begin{eqnarray}\label{eqn:mxcjpaon}
 T_{\rm sys,on,mxc} &=& \frac{2T_{\rm mxc}+T_{{\rm JPA}}}{\alpha_1}\nonumber \\
    & & + \frac{T_{\rm mxc}(1-\alpha_2)+T_{\rm HFET}}{\alpha_1\alpha_2 G_{\rm JPA}}. 
\end{eqnarray}
Since the recorded physical temperature is $T_{\rm mxc}$ in Eq.~\ref{eqn:mxcjpaoff} and Eq.~\ref{eqn:mxcjpaon}, the information from cavity cool-down or warm-up data is the part without the transmissivity ($\alpha_1$, $\alpha_2$), i.e.~$T_{\rm HFET}$ and $T_{\rm JPA}$, which can provide extra understanding of the system while we know the transmissivities ahead of time.

However, it's common to have the temperature gradients $\mathcal{O}(0.1T_{\rm mxc})$ among the components that are supposed to be well-thermalized to $T_{\rm mxc}$. For the JPA-off case, Eq.~\ref{eqn:mxcjpaoff} is still practical especially using the cavity off-resonance data because $T_{\rm HFET}$ is often more than an order of magnitude larger than the other noise contributions in Eq.~\ref{eqn:tsysOff}. For the JPA-on case, the temperature gradients $\mathcal{O}(0.1T_{\rm mxc})$ are so large that Eq.~\ref{eqn:mxcjpaon} fails the ideal assumption, and Eq.~\ref{eqn:tsysOn} is used instead.

\section{Noise Calibration Techniques}\label{sec:measurements}
In this section, we present the noise calibration techniques with examples from the most recent ADMX data taking run during 2024~\cite{run1D_TBP}. Compared to previous ADMX runs~\cite{admx2020,admx2021,admx2024}, a stronger thermal link is used to connect the hot load to the milliKelvin space ($80\sim100$~mK). The base temperature of the hot load reaches $140\sim170$~mK, which is cold enough to perform the JPA-on-hot-load noise measurement without saturating the JPA. All the noise calibrations under different circumstances are Y-factor measurements where the output powers are traced as a function of the physical temperatures, and a linear-fit is used to extract out the extra electronic noises introduced by the different components in the receiver chain. More specifically, for the JPA-off measurements the fit function is of the form

\begin{equation}\label{eqn:fit_fcn_jpa_off}
    P_{\rm off} = C(T + T_{\rm fit}).
\end{equation}
Here, $P_{\rm off}$ is the output power with the JPA off, $T$ is the temperature that is being changed, $T_{\rm fit}$ is the noise temperature we fit out for each measurement and C is a constant. 

For the JPA-on measurements, we need to correct for JPA gain due to inevitable fluctuations during the course of the measurement. To do this, we change the left hand side of the fit function from $P_{\rm on}$ to $(P_{\rm on} - P_{\rm off})/G_{\rm JPA}$. We see that 

\begin{eqnarray}  
\label{eqn:gain_corrected_Tsys_full}
    & &\frac{P_{\rm on} - P_{\rm off}}{G_{\rm JPA}} \propto  T_{\rm sys,on} - 
\frac{T_{\rm sys,off}}{G_{\rm JPA}} =  \nonumber\\
    && ({T_{\rm stage1}} |\Gamma_{\rm cav}|^2  + T_{\rm cav} (1-|\Gamma_{\rm cav}|^2))(1-\frac{1}{G_{\rm JPA}}) \nonumber\\ 
   & &+ {T_{\rm stage1}} + 2\frac{T_{\rm circ}(1-\alpha_1)}{\alpha_1}+
    \frac{T_{{\rm JPA}}}{\alpha_1} \nonumber \\
    & &- \frac{T_{\rm circ}(1-\alpha_1)}{\alpha_1 G_{\rm JPA}}. 
\end{eqnarray}

For sufficiently high gain $(1-\frac{1}{G_{\rm JPA}} \simeq 1$ and $\frac{T_{\rm circ}(1-\alpha_1)}{\alpha_1 G_{\rm JPA}} \simeq 0$), the model we use for gain-corrected measurements is

\begin{eqnarray}   
\label{eqn:gain_corrected_Tsys_simplified}
   T_{\rm sys,on} - 
\frac{T_{\rm sys,off}}{G_{\rm JPA}} \simeq & & {T_{\rm stage1}} |\Gamma_{\rm cav}|^2  + T_{\rm cav} (1-|\Gamma_{\rm cav}|^2) \nonumber\\ 
& &+ {T_{\rm stage1}} + 2\frac{T_{\rm circ}(1-\alpha_1)}{\alpha_1} \nonumber \\
&&+\frac{T_{{\rm JPA}}}{\alpha_1}, 
\end{eqnarray}
which is the same as Eq.~\ref{eqn:tsysOn} without the final two terms. 

That being said, the form of the fit function for the JPA-on measurements is nearly identical to the JPA-off function, aside from the gain correction to the output power on the left hand side

\begin{equation}\label{eqn:fit_fcn_jpa_on}
    \frac{P_{\rm on} - P_{\rm off}}{G_{\rm JPA}} = C(T + T_{\rm fit}).
\end{equation}
The errors reported for the individual fit results in the following subsections are purely statistical to reflect the quality of our data and, as such, do not include systematic error. The primary source of systematic error in each fitting is calibrated ruthenium oxide temperature sensors which have a known offset of about $\pm 5$ mK. We also consider the systematic uncertainties introduced by the choice of slightly different temperature windows during fitting when reporting the final values as shown in Tab.~\ref{tab:comparison}. 

In the rest of this section, we present the details of the noise calibration measurements with the cryo-switch flipped to the hot load or the cavity system and with JPA unpowered or powered in sequence, and we further compare the system noise with a direct JPA-on noise measurement to the SNRI method. For all measurements done with the JPA powered on, the JPA bias settings (bias current and pump power) were optimized at the start of the measurement to achieve the highest possible gain with the highest possible stability. Stability was prioritized over magnitude, as we did not rebias the JPA throughout the course of the measurements due to the time-intensive nature of the process. Therefore, it was paramount that the gain remain as stable as possible in order to get the cleanest fits to the data. As mentioned earlier, the pump tone was centered 320 kHz higher than the nominal resonant frequency of the cavity ($f_0$), ensuring that the $f_0$ resides comfortably within the JPA's bandwidth while not interfering with the cavity resonance.

\subsection{JPA Off Hot Load}\label{sec:JPAoffhotload}
The JPA-off-hot-load measurement is performed by powering down the JPA, so we can calibrate the noise coming from the second stage HFET amplifier. The relevant model in this instance is Eq.~\ref{eqn:hljpaoff}. The fit function used for this measurement is Eq.~\ref{eqn:fit_fcn_jpa_off}. More specifically, $T = T_{\rm HL}$ , and $T_{\rm fit} =\frac{T_{\rm circ}(1-\alpha)}{\alpha}  +\frac{T_{\rm HFET}}{\alpha}$ where $\alpha=\alpha_1\alpha_2$. We refer to this fit result as the \textit{effective} HFET noise, ${T_{\rm HFET}}/{\alpha}_{\rm eff}$, due to the inclusion of the circulator term and the scaling of $1/\alpha$, whereas the intrinsic HFET noise is equal to $T_{\rm HFET}$.

\begin{figure}[h]
    \includegraphics[width=\linewidth]{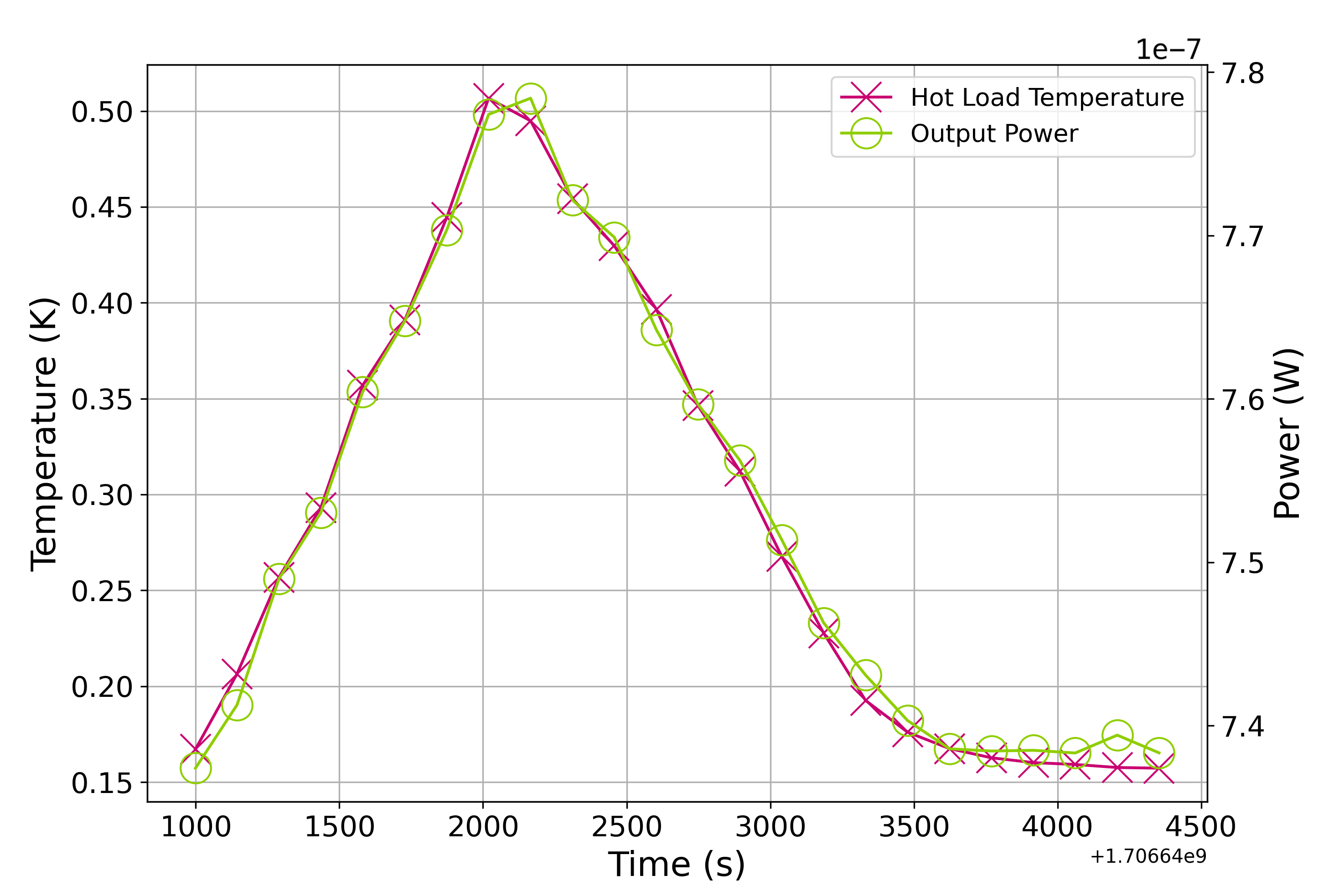}
    \caption{HFET hot load data at 1280 MHz. Here we plot the temperature of the hot load ($T_{\rm HL}$) and the output power versus time respectively as an example of what the raw data looks like. Over the course of the measurement as the load is heated up and cooled down the output power rises and falls correspondingly. Using this relationship we can fit the two quantities, power and $T_{\rm HL}$, against each other to extract the effective HFET noise, $T_{\rm HFET}/\alpha_{\rm eff}$. The fit of this data can be seen in Fig~\ref{fig:HFET_HL_1280}. Temperature and power data taken over time, shown here as an example, are also used to produce the fits shown in Figs.~\ref{fig:HFET_Cavity_1280},\ref{fig:JPA_HL_1280},\ref{fig:JPA_cavity_1280_offres},and \ref{fig:JPA_cavity_1280_onres}.}
    \label{fig:HFET_HL_1280_vs_time}
\end{figure}

\begin{figure}[h]
    \includegraphics[width=\linewidth]{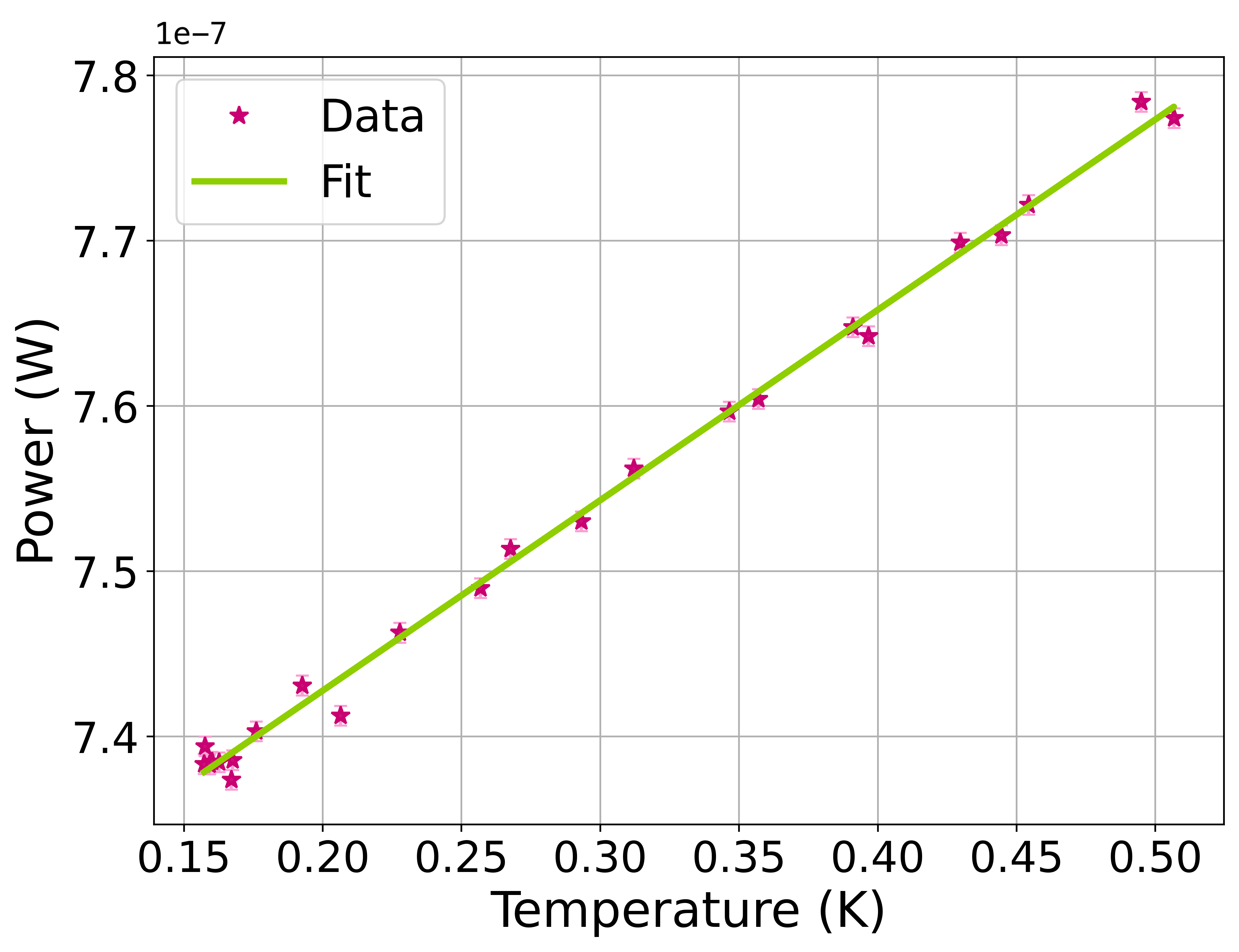}
    \caption{HFET hot load measurement at 1280 MHz. Here we plot the output power versus the temperature of the hot load ($T_{\rm HL}$) to fit out $T_{\rm HFET}/\alpha_{\rm eff}$. The fit shown was done using the entire data range giving $T_{\rm HFET}/\alpha_{\rm eff} = 6.25 \pm 0.09$ K when looking at the hot load at 1280 MHz. After averaging the fitting results obtained by using different temperature windows, we find $T_{\rm HFET}/\alpha_{\rm eff} = 6.13 \pm 0.20$ K.}
    \label{fig:HFET_HL_1280}
\end{figure}

The procedure is as follows. We begin by flipping the RF switch in Fig.~\ref{fig:rf_layout} from the cavity position to the hot load position. Then, we connect the hot load to a DC power supply, and begin adding heat incrementally allowing for the temperature to settle at each stage before moving on. Due to the broadband coverage of the HFET, we are able to digitize the power at multiple frequencies during this measurement. We typically do about 10-20 frequency points per measurement, spaced a few MHz apart. After heating the load to roughly 0.5-1 K, we begin ramping the heater down, continuing to measure output power until we return to the base temperature of the hot load ($\sim150$ mK). Figure \ref{fig:HFET_HL_1280_vs_time} provides an example of this type of measurement at 1280~MHz where we track the output power and the temperature of the hot load at the same time. The fit of Eq.~\ref{eqn:fit_fcn_jpa_off} with this data resulting in ${T_{\rm HFET}}/{\alpha}_{\rm eff}=6.13\pm 0.20$~K can be seen in Fig.~\ref{fig:HFET_HL_1280}.

\subsection{JPA Off Cavity Cool-Down/Warm-up}\label{sec:jpaoffcooldown}
The JPA-off-cavity measurement is also performed with the JPA powered off, but the RF switch in Fig.~\ref{fig:rf_layout} is flipped to the cavity position. Compared to Sec.~\ref{sec:JPAoffhotload}, the cavity cannot be heated or cooled in isolation like the hot load can, so the entire system is either cooling down or warming up together. Therefore, Eq.~\ref{eqn:mxcjpaoff} is the relevant model for this case. We still use Eq.~\ref{eqn:fit_fcn_jpa_off} as the fit function, but now $T = T_{\rm mxc}$ ($T = T_{\rm circ}$) and $T_{\rm fit} = T_{\rm HFET}$. As one can see, this measurement allows us to fit out $T_{\rm HFET}$, without the factor of ${1}/{\alpha}$ present in the hot load measurement. 

\begin{figure}[h]
    \includegraphics[width=\linewidth]{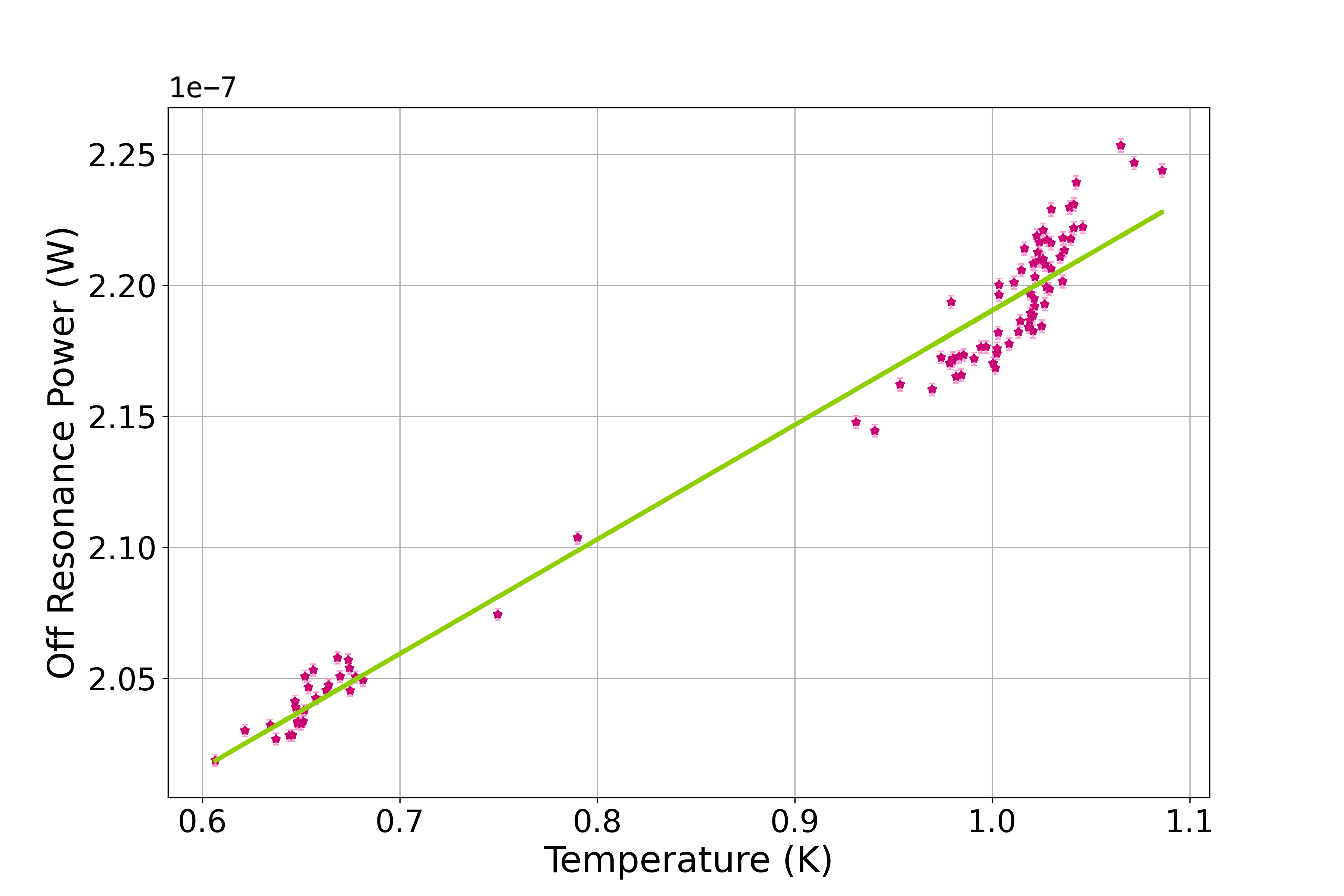}
    \caption{HFET cavity cool-down measurement at 1280 MHz. Here we plot the output power versus $T_{\rm mxc}$ from Eq.~\ref{eqn:mxcjpaoff} to fit out $T_{\rm HFET}$. This fit was done using the full data range and for this scenario we find that $T_{\rm HFET} = 4.02\pm0.08$~K. The assumption in Eq.~\ref{eqn:mxcjpaoff} that $T_{\rm mxc} = T_{\rm circ}$ is not perfect, so we perform the fit a second time on the full data range using $T_{\rm circ}$ instead of $T_{\rm mxc}$ which gives $T_{\rm HFET} = 3.76\pm0.09$~K. The average of all the fitted values for $T_{\rm HFET}$ using different temperature windows for both the $T_{\rm mxc}$ and $T_{\rm circ}$ cases is $4.18 \pm 0.26$~K.}
    \label{fig:HFET_Cavity_1280}
\end{figure}

With the JPA-off-cavity measurement, we can do two additional diagnostics that can help characterize our RF chain. Firstly, we can compare the measured value of $T_{\rm HFET}$ directly to the data sheet to ensure it is working as expected. Secondly, we can combine this result with the hot load result to back out the total transmissivity, $\alpha$, and compare to measurements of $\alpha$ done before data taking. We are able to do this with the knowledge that the magnitude of the circulator term is less than 1\% of the magnitude of the HFET term in Eq.~\ref{eqn:hljpaoff}, so $T_{\rm HFET}/\alpha_{\rm eff} \simeq T_{\rm HFET}/\alpha$.
This simplified model requires the assumption that, on resonance, the cavity and circulators are all well thermalized to the mixing chamber, and off resonance, attenuator A and circulators are all well thermalized to the mixing chamber. We find that this assumption is more true in the off resonance case as the cavity thermalizes more slowly than the other components. Therefore, we only use the off resonance data for this analysis so we can get the most accurate measurement of $T_{\rm HFET}$ and thus the most accurate measurement of $\alpha$ when combined with $T_{\rm HFET}/\alpha_{\rm eff}$ from the JPA-off-hot-load measurement. Figure \ref{fig:HFET_Cavity_1280} shows an example of this type of measurement at 1280~MHz giving $T_{\rm HFET}=4.18 \pm 0.26$~K, which is reasonable according to the HFET calibration data from Low Noise Factory~\cite{LNF}. After combining this result with that shown in Fig.~\ref{fig:HFET_HL_1280}, $\alpha = 0.68\pm 0.05$, which agrees with the insertion loss measured before data taking $\alpha=0.643 \pm 0.003$. The pre-data-taking insertion loss measurement was of $\alpha_1$, the cavity-to-JPA insertion loss. We inferred the total insertion loss, $\alpha$, by assuming it is dominated by the identical circulators such that $\alpha=\alpha_1^2$ (i.e. $\alpha_1$ = $\alpha_2$). 

\subsection{JPA On Hot Load}\label{sec:JPAonhotload}
The total gain is less stable with the JPA on because the JPA is a narrowband amplifier, unlike the HFET and warm (post) amplifiers, so slight changes in its environment such as temperature fluctuations or mechanical vibrations can be enough to alter its optimal bias parameters and change the gain. Therefore, the JPA-on-hot-load measurement is similar to the JPA-off-hot-load measurement, but requires a few more steps because of the decreased gain stability. Additionally, the model is more complex in this case (see Eq.~\ref{eqn:gain_corrected_Tsys_simplified}).
More specifically, we fit the gain corrected power, $(P_{\rm on}-P_{\rm off})/G_{\rm JPA}$, using Eq.~\ref{eqn:fit_fcn_jpa_on}, where $T = 2T_{\rm HL}$ and $T_{\rm fit} =2\frac{T_{\rm circ}(1-\alpha_1)}{\alpha_1} + \frac{T_{{\rm JPA}}}{\alpha_1}$. The factor of 2 in the definition of $T$ comes from the addition of the idler mode noise power when the JPA is on. We refer to this fit result as the effective JPA noise, $T_{\rm JPA,eff}$. Here, $T_{\rm JPA}$ is the intrinsic excess noise from the JPA as defined in Sec.~\ref{Example Haloscope Model}. The circulator term is expected to contribute on the order of 50~mK worth of noise to $T_{\rm JPA,eff}$ in our system which is not negligible when compared to the $T_{\rm JPA}$ term, so we are careful to call this the \textit{effective} JPA noise. 

As previously mentioned, the procedure is nearly identical to the JPA off hot load measurement with a few additional steps. We again begin by flipping the RF switch in Fig.~\ref{fig:rf_layout} from the cavity position to the hot load position. This introduces some heat to the JPA, which can be very sensitive to changes in temperature. Therefore, we wait a few minutes for the JPA temperature to level out before we begin attempting to adjust the JPA DC bias current and pump power to get a decent gain. Once we are satisfied with the magnitude of the JPA gain, we then test to make sure the gain is stable with the given settings. We vary the bias current and pump power over a small range and monitor how much the gain changes. If it fluctuates too much, we repeat the process of manually adjusting the parameters and look for a new gain point to test. If the gain appears fairly stable, we begin monitoring the gain over time before adding heat to the system to further test for stability. Once the gain remains stable, we connect the hot load to a DC power supply, and begin adding heat incrementally. 

\begin{figure}[h]
    \includegraphics[width=\linewidth]{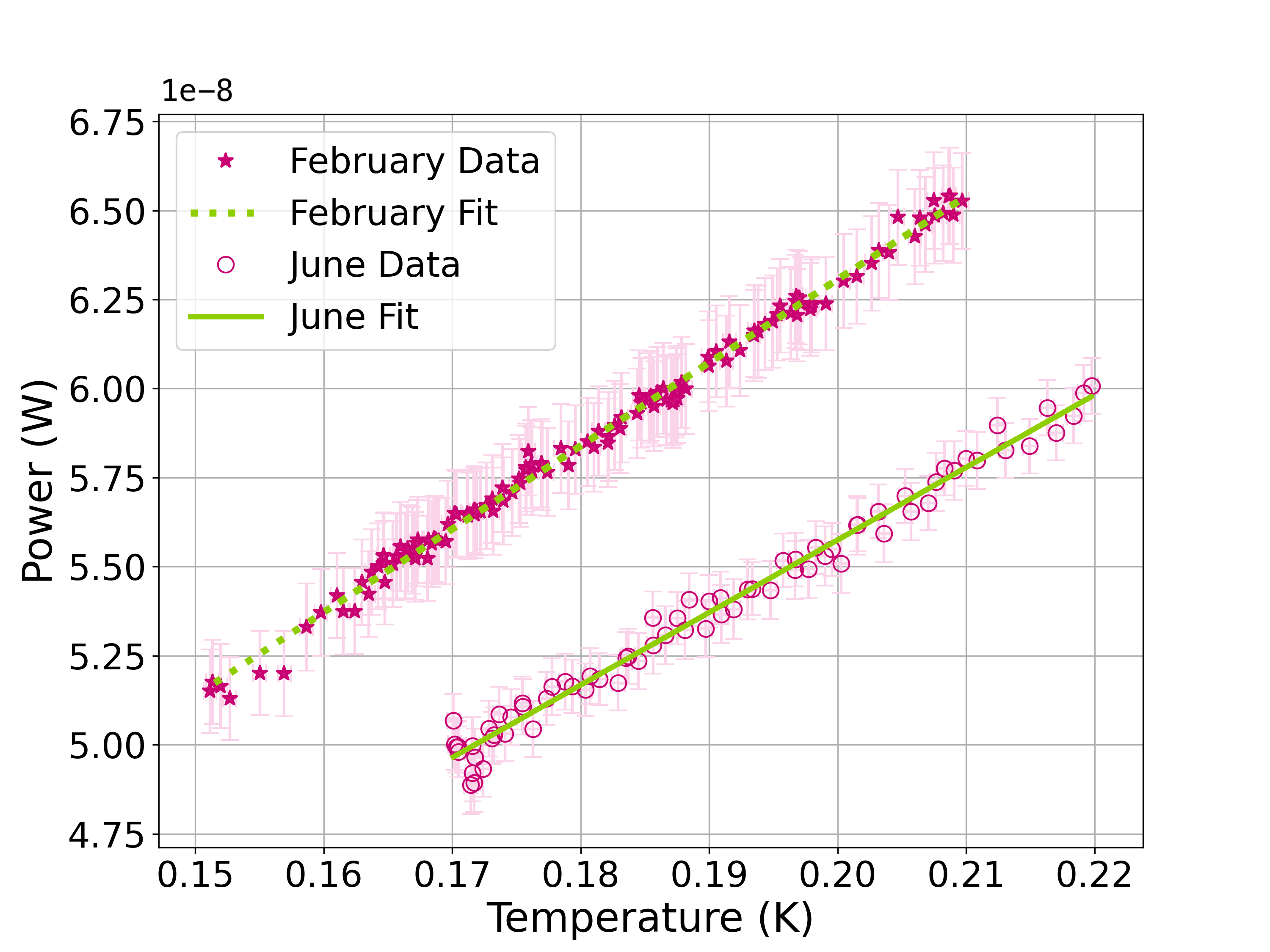}
    \caption{JPA hot load measurements at 1280~MHz. Here we plot the output power versus the temperature of the hot load for both the February and June measurements to fit out the effective JPA noise ($T_{\rm JPA,eff} = 2\frac{T_{\rm circ}(1-\alpha_1)}{\alpha_1} + \frac{T_{{\rm JPA}}}{\alpha_1}$). The two fits shown here were done using the entire data range, and for this scenario we find that $T_{\rm JPA,eff,February} = 0.140 \pm 0.004$~K and $T_{\rm JPA,eff,June} = 0.148 \pm 0.008$~K when looking at the hot load at 1280 MHz. These results are consistent within 1$\sigma$ which gives us confidence that at a given frequency the effective JPA noise is stable on the timescale of a few months. Taking different temperature fitting windows into account, the averaged effective JPA noises are $T_{\rm JPA,eff,February} = 0.139 \pm 0.021$~K and $T_{\rm JPA,eff,June} = 0.143 \pm 0.019$~K which are still consistent with each other indicating stability with different JPA bias settings over time (see text for specific values of $I_{\rm bias}$ and $P_{\rm pump}$).}
    \label{fig:JPA_HL_1280}
\end{figure}

Unlike the HFET, the JPA is a narrow-band amplifier, so we perform this measurement at one frequency at a time, continuously measuring the gain and output power at the target frequency. After heating the load to a maximum temperature of roughly 200~mK, we begin ramping the heater down, continuing to measure output power until we return to the base temperature ($\sim150$~mK). We do not take the hot load much higher than 200~mK with the JPA on because it can quickly become saturated and/or lose gain performance. We performed this measurement twice at 1280~MHz, with a difference of about 4 months between the two measurements to test for stability of the effective JPA noise. Both measurements done at 1280~MHz can be seen in Fig~\ref{fig:JPA_HL_1280}. The two measurements were done with different gains: $G_{\rm JPA,February}=15.8 \pm 0.1 $~dB ($I_{\rm bias}=-0.183 \rm~mA$ and $P_{\rm pump}=-7.35 \rm~dBm$) and $G_{\rm JPA,June}=18.1 \pm 0.8$~dB ($I_{\rm bias}=-1.647 \rm~mA$ and $P_{\rm pump}=-8.47 \rm~dBm$), which share consistent $T_{\rm JPA,eff}$: $T_{\rm JPA,eff,February}=0.139\pm 0.021$~K and $T_{\rm JPA,eff,June}=0.143\pm 0.019$~K. Note that the pump powers we reported are not the absolute powers at the JPA reference plane but the output of the signal generator at room temperature.

\subsection{JPA On Cavity Cool-Down/Warm-up}\label{sec:JPACavityCooldown}

Similar to the measurement described in Sec.~\ref{sec:jpaoffcooldown}, the JPA-on-cavity measurement takes place with the RF switch in Fig.~\ref{fig:rf_layout} flipped to the cavity position, but this time the JPA is powered on. As discussed previously, during this measurement the entire system is either cooling down or warming up together. Data from this measurement should follow Eq.~\ref{eqn:mxcjpaon}. However, as described in Sec.~\ref{sec:jpaoffcooldown}, Eq.~\ref{eqn:mxcjpaon} requires good thermalization among different milliKelvin electronics to extract out $T_{\rm JPA}$, which is not practical with $\mathcal{O}(0.1T_{\rm mxc})$ temperature gradients.
Therefore, we leave Eq.~\ref{eqn:mxcjpaon} as a theoretical framework which would be applicable in the case that our system was exceptionally well thermalized, and use the full model described in Eq.~\ref{eqn:tsysOn}. 

Since the reflectivity of the cavity is significantly different between the off and on resonance, we separate the fittings accordingly and take into account the different temperatures of individual components as well as the reflectivity of the cavity. The gain of the JPA during the course of this measurement varied from roughly 11-19.5 dB with the JPA bias settings kept constant ($I_{\rm bias}=-1.038 \rm~mA$ and $P_{\rm pump}=-6.46 \rm~dBm$) as the system heated up. As a result, the analysis of this data required the power to be corrected for gain fluctuations as was done for the JPA-on-hot-load measurements. 

The fit function used for the JPA-on-cavity is Eq.~\ref{eqn:fit_fcn_jpa_on}. For the off-resonance data, $T=2T_{\rm stage1} = 2(T_{A} 
\sqrt{\alpha_1}+ T_{\rm{circ}} (1-\sqrt{\alpha_1}))$ and $T_{\rm fit} = T_{\rm JPA,eff}$ as defined in Sec.~\ref{sec:admxModel} and Sec.~\ref{sec:JPAonhotload} respectively. As for the JPA-on-hot-load fit, the factor of 2 in the definition of $T$ is due to the addition of idler mode noise. Note that in the above definition of $T_{\rm stage1}$ we make the assumption that the attenuation between attenuator A and the cavity is equal to half the attenuation from the cavity to JPA ($\alpha_{\rm{circ1},21} = \sqrt{\alpha_1}$). For the on-resonance data, $T = T_{\rm stage1}  |\Gamma_{\rm cav}|^{2} + T_{\rm cav}  (1-|\Gamma_{\rm cav}|^{2}) + T_{\rm stage1} $ and $T_{\rm fit} = T_{\rm JPA,eff}$. Here, the temperatures of the signal and idler modes are too different to combine into a single term so we use the full definition for $T$.

\begin{figure}[h]
    \includegraphics[width=\linewidth]{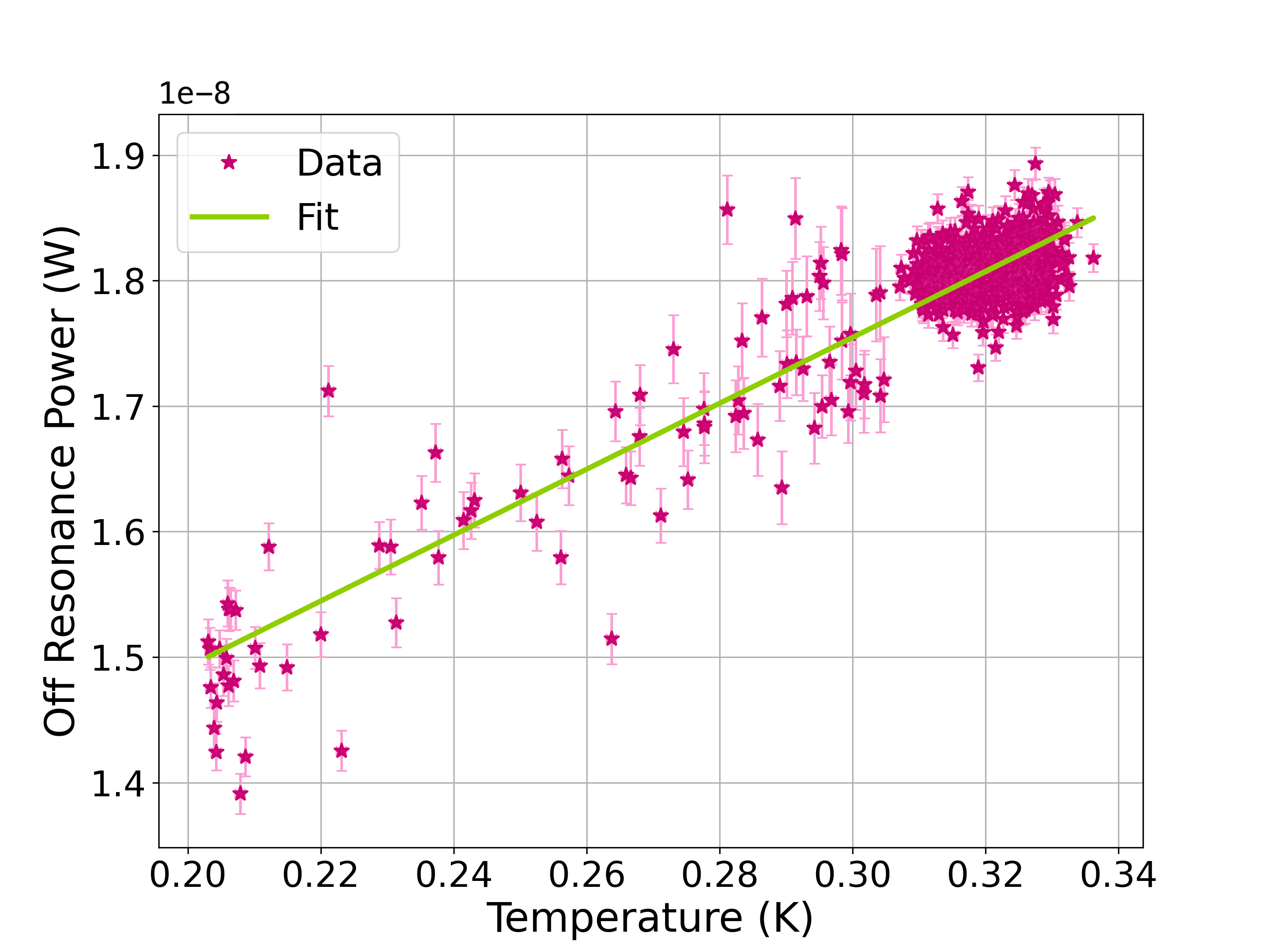}
    \caption{Off resonance JPA-on-cavity warm-up measurement at 1280 MHz. Here we plot the gain corrected off resonance output power versus $2T_{\rm stage1}$ to fit out the effective JPA noise ($T_{\rm JPA,eff} = 2\frac{T_{\rm circ}(1-\alpha_1)}{\alpha_1} + \frac{T_{{\rm JPA}}}{\alpha_1}$). The entire data range fit results in $T_{\rm JPA,eff} = 0.368 \pm 0.014$~K. The average of the fit results obtained using different data ranges is $T_{\rm JPA,eff} = 0.372 \pm 0.022$~K.}
    \label{fig:JPA_cavity_1280_offres}
\end{figure}

\begin{figure}[h]
    \includegraphics[width=\linewidth]{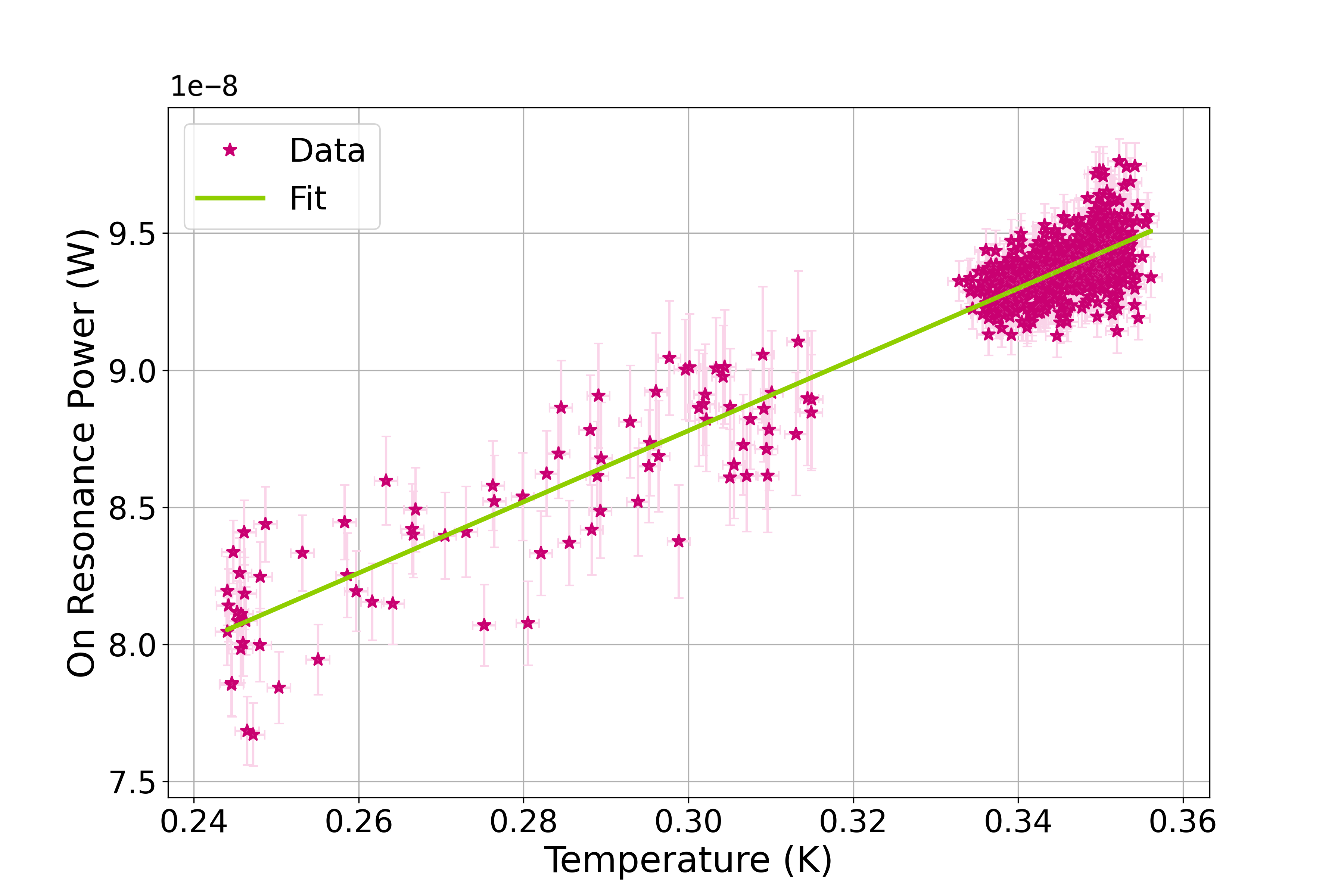}
    \caption{On resonance JPA-on-cavity warm-up measurement at 1280 MHz. Here we plot the gain corrected on resonance output power versus $T_{\rm stage1}(1+|\Gamma_{\rm cav}|^2)  + T_{\rm cav} (1-|\Gamma_{\rm cav}|^2)$ to fit out the effective JPA noise ($T_{\rm JPA,eff} = 2\frac{T_{\rm circ}(1-\alpha_1)}{\alpha_1} + \frac{T_{{\rm JPA}}}{\alpha_1}$). The fit shown in this plot was done using the entire data range, which results in $T_{\rm JPA,eff} = 0.377 \pm 0.013$~K. The average of the fit results obtained using different data ranges is $T_{\rm JPA,eff}=0.372 \pm 0.018$~K.}
    \label{fig:JPA_cavity_1280_onres}
\end{figure}

An example of off~(on) resonance JPA-on-cavity measurement can be seen in Fig.~\ref{fig:JPA_cavity_1280_offres}~(Fig.~\ref{fig:JPA_cavity_1280_onres}) at 1280~MHz, giving $T_{\rm JPA,eff}=0.372\pm0.022$~K~($0.372\pm0.018$~K), which is mysteriously higher than the JPA-on-hot-load $T_{\rm JPA,eff}$~(Fig.~\ref{fig:JPA_HL_1280}). Possible reasons for this discrepancy are discussed in Section~\ref{sec:conclusion}.

\subsection{System Noise Temperature Comparison}

Before we were able to perform a direct JPA on noise measurement, we calculated our system noise temperature ($T_{\rm sys}$) in two steps. The first step would be to directly measure the JPA off noise $T_{\rm HFET}/\alpha_{\rm eff}$, as described in Sec.~\ref{sec:JPAoffhotload}. Then, we measure the signal-to-noise-ratio improvement (SNRI) as defined in Eq.~\ref{eqn:SNRI}. We can then combine the results of these two measurements to calculate our system noise temperature using Eq.~\ref{eqn:SNRI_tsys}. 

\begin{figure}[h]
    \includegraphics[width=\linewidth]{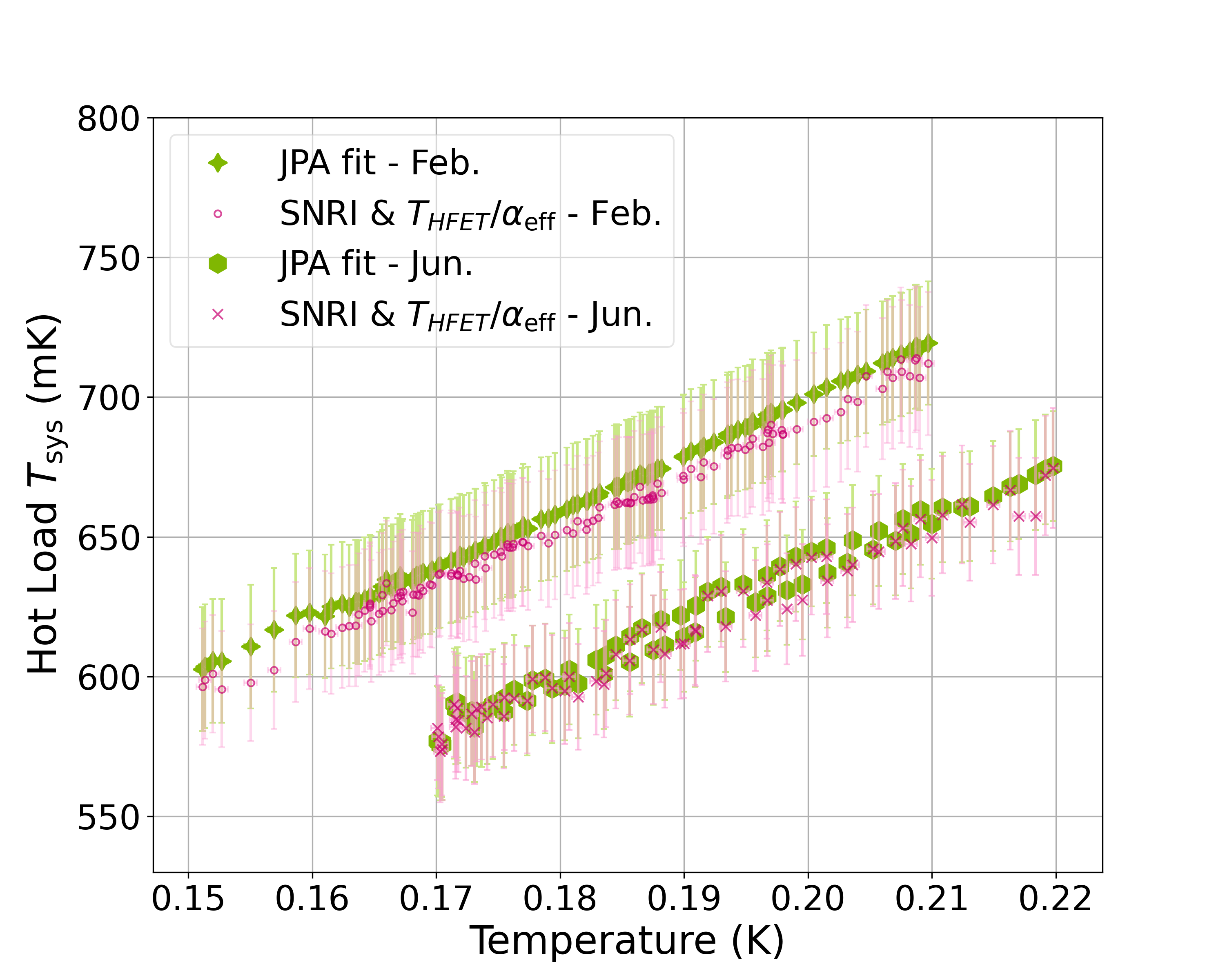}
    \caption{Hot Load $T_{\rm sys}$ comparison at 1280 MHz. Here we show the comparison between the two methods, using SNRI (pink points) and using JPA fit (green points), over the course of both hot load measurements done at 1280 MHz. It is clear that for both the February and June data the system noise temperature we calculate is consistent between the two methods within error bars.}
    \label{fig:1280_Tsys_comp_HL}
\end{figure}

\begin{figure}[h]
    \includegraphics[width=\linewidth]{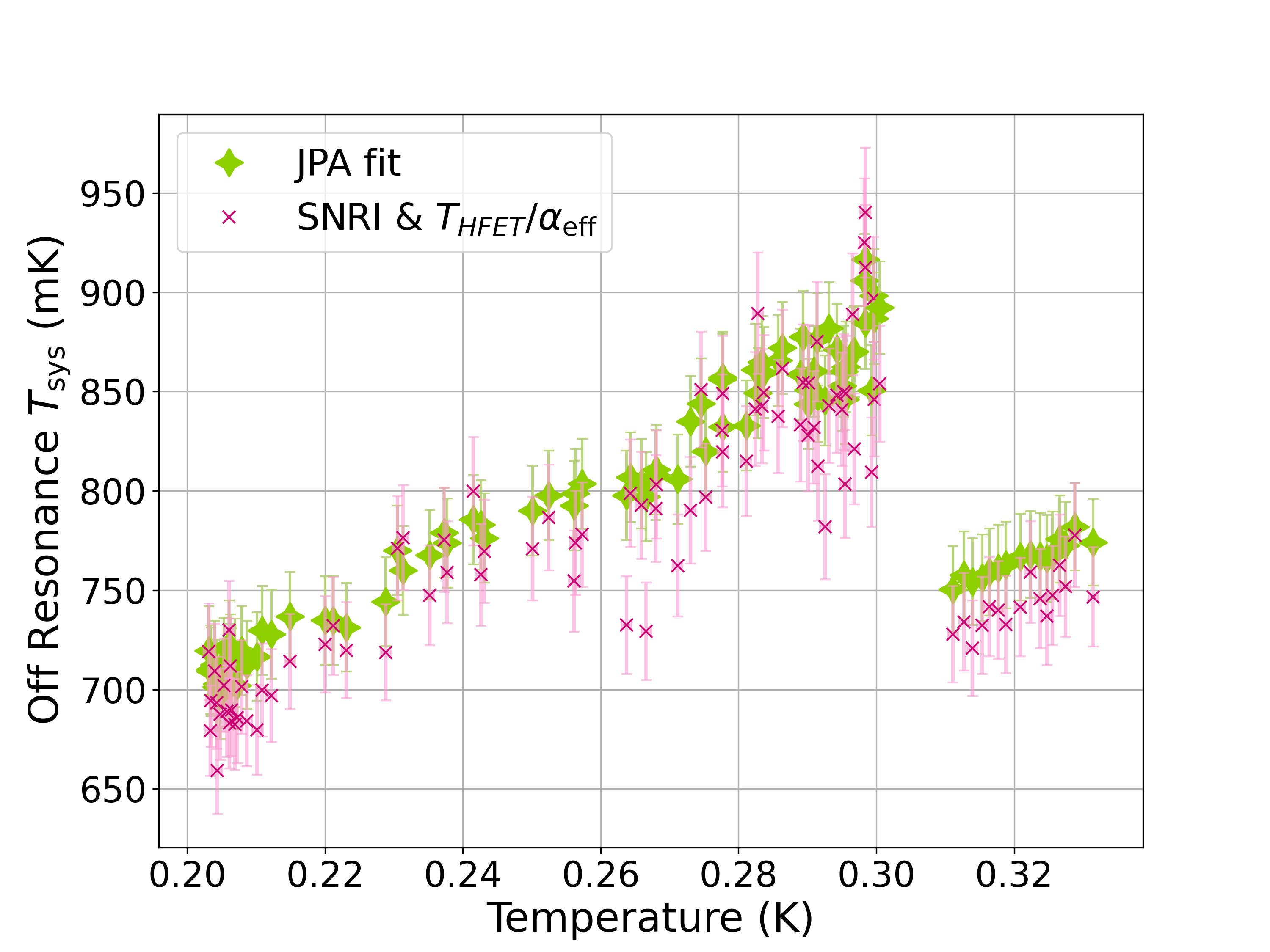}
    \caption{Off resonance cavity $T_{\rm sys}$ comparison at 1280 MHz.  Here we show the comparison between the two methods, using SNRI (pink points) and using JPA fit (green points), over the course of the cavity cool-down measurement done at 1280 MHz (off resonance data only). Data above 0.3~K have been randomly downsampled for plotting purposes due to the high density of data in that region. The discontinuity around this temperature was caused by a sharp increase in JPA gain during the course of the measurement, causing a drop in $T_{\rm sys}$. There is about a 21 mK difference between the two methods on average, but taking into account the error bars, the two methods can be considered consistent.}
    \label{fig:1280_Tsys_comp_cav_offres}
\end{figure}

\begin{figure}[h]
    \includegraphics[width=\linewidth]{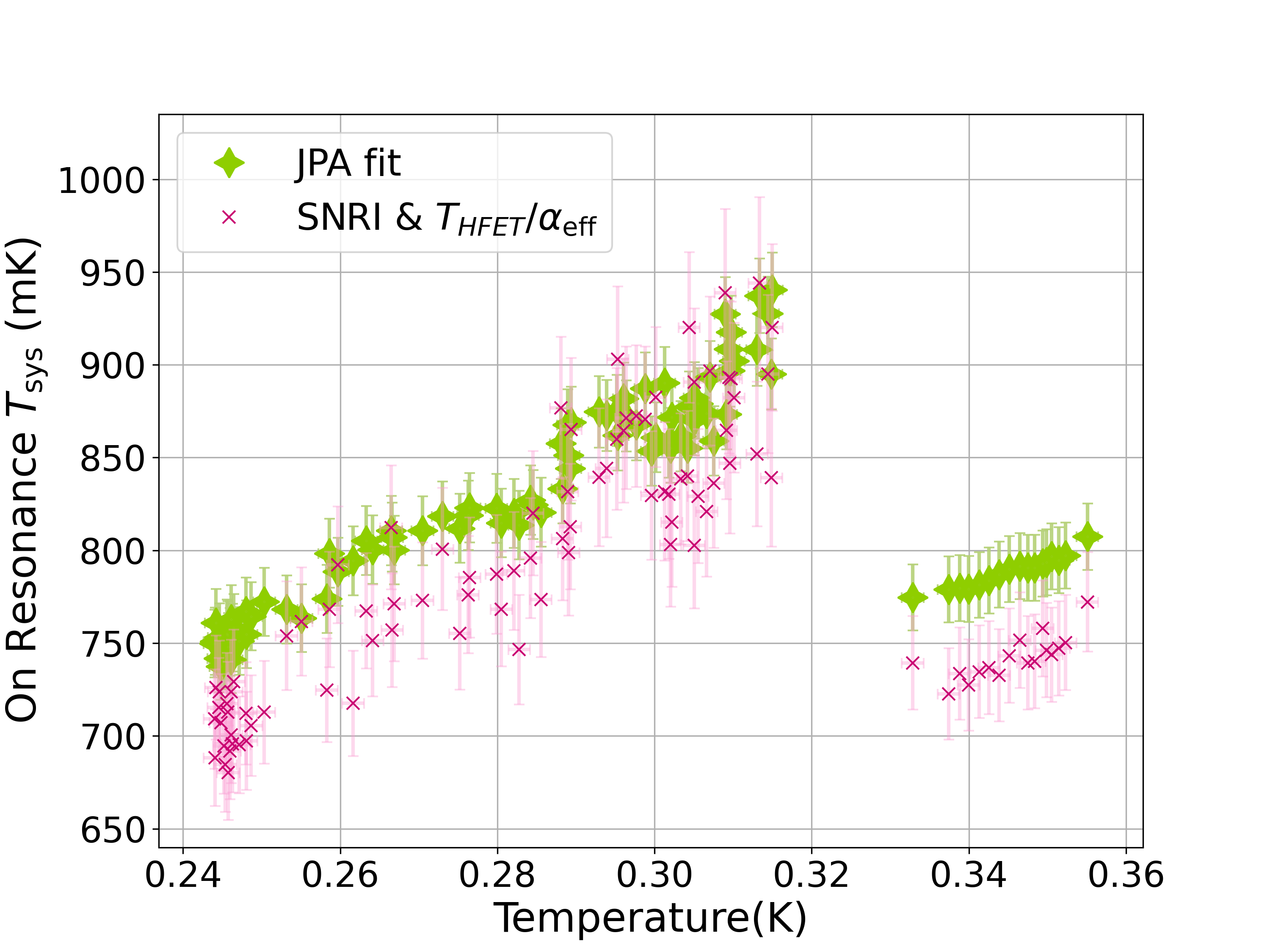}
    \caption{On resonance $T_{\rm sys}$ comparison at 1280 MHz. Here we show the comparison between the two methods, using SNRI (pink points) and using JPA fit (green points), over the course of the cavity cool-down measurement done at 1280 MHz (on resonance data only). Data above 0.32~K have been randomly downsampled for plotting purposes due to the high density of data in that region. The discontinuity is explained in Fig.~\ref{fig:1280_Tsys_comp_cav_offres}. There is about a 45 mK difference between the two methods on average, which corresponds to $1.3\sigma$ difference.
}
    \label{fig:1280_Tsys_comp_cav_onres}
\end{figure}

Now with the ability to measure the JPA effective noise, $T_{\rm JPA,eff}$, directly, we can use the full model for $T_{\rm sys}$ defined in Eq.~\ref{eqn:tsysOn}. In this section, we compare the two methods for calculating our system noise temperature.
An example showing the SNRI and direct JPA-on fit methods for the hot load measurement at 1280 MHz can be seen in Fig.~\ref{fig:1280_Tsys_comp_HL}. Similarly, a comparison between the SNRI and direct JPA fit methods for the off (on) resonance cavity measurement at 1280 MHz can be seen in Fig.~\ref{fig:1280_Tsys_comp_cav_offres} (Fig.~\ref{fig:1280_Tsys_comp_cav_onres}). 
We see that the results for the system noise temperature, $T_{\rm sys}$, are consistent between the two methods for both the hot load and the off resonance cavity measurement, and nearly consistent for the on resonance cavity measurement as well. This gives us confidence that the model in Eq.~\ref{eqn:tsysOn} effectively describes how $T_{\rm sys}$ depends on the various parameters in the receiver chain. Additionally, the slight differences between the two methods provide us with an estimate of the systematic uncertainty in $T_{\rm sys}$ to use for the axion search analysis for this data taking run.

To trace $T_{\rm sys}$ during data taking where the cryo-switch is flipped to the cavity, we resort to the SNRI method for its promptness with $T_{\rm HFET}/\alpha_{\rm eff}$ as a calibrated input. The JPA-off-hot-load measurement fitting directly provides $T_{\rm HFET}/\alpha_{\rm eff}$~(Sec.~\ref{sec:JPAoffhotload}). The JPA-on-hot-load and JPA-on-cavity measurements need further calculation where
\begin{eqnarray}\label{eqn:inferredThfetOverAlpha}
   T_{\rm HFET}/\alpha_{\rm eff} &=&  \frac{1 }{G_{\rm JPA} -{\rm SNRI}}\cdot \nonumber \\
& & \bigg( {\rm SNRI}\cdot G_{\rm JPA} \cdot (T_{\rm JPA,eff}+T_{\rm JPA,on}) \nonumber\\
 & & - G_{\rm JPA} \cdot T_{\rm JPA,off}  \nonumber\\
 & & - {\rm SNRI} \cdot T_{\rm circ} (1-\alpha_2)/\alpha_2 \bigg) .
\end{eqnarray}

In the JPA-on-hot-load case, $T_{\rm JPA,on} =2 T_{\rm HL}$ and $T_{\rm JPA,off} =T_{\rm HL}$. In the JPA-on-cavity case, $T_{\rm JPA,on} =T_{\rm stage1}  |\Gamma_{\rm cav}|^{2} + T_{\rm cav}  (1-|\Gamma_{\rm cav}|^{2}) + T_{\rm stage1}$ and $T_{\rm JPA,off} =T_{\rm stage1}  |\Gamma_{\rm cav}|^{2} + T_{\rm cav}  (1-|\Gamma_{\rm cav}|^{2} )$. For the off-resonance JPA-on-cavity case it is assumed that $|\Gamma_{\rm cav}|^{2} = 1$. 
Putting the $T_{\rm JPA,eff}$ calibration results at 1280~MHz reported in Sec.~\ref{sec:JPAonhotload} into the equation above, $T_{\rm HFET}/\alpha_{\rm eff}=6.18 \pm 0.21 $~K after averaging the two measurements separated by four months. Similarly, putting the $T_{\rm JPA,eff}$ calibration results at 1280~MHz reported in Sec.~\ref{sec:JPACavityCooldown} into the equation above, $T_{\rm HFET}/\alpha_{\rm eff}=6.72 \pm 0.24 ~(6.33 \pm 0.25)$~K on (off) resonance. The inferred $T_{\rm HFET}/\alpha_{\rm eff}$ values from the JPA-on-hot-load and JPA-on-cavity off resonance calibrations are consistent with the JPA-off-hot-load result within $1\sigma$, $T_{\rm HFET}/\alpha_{\rm eff} = 6.13\pm 0.21$~K. The JPA-on-cavity on resonance inferred $T_{\rm HFET}/\alpha_{\rm eff}$ is nearly consistent ($< 1.4\sigma$) with the JPA-off-hot-load result as well.

\section{Discussion and conclusion}\label{sec:conclusion}
 Noise calibration results in different conditions are summarized in Tab.~\ref{tab:comparison} with the examples at 1280~MHz for the ADMX haloscope. Additionally, the physical temperatures of relevant components during each of the JPA-on measurements are summarized in Tab.~\ref{tab:temperatures}. Comparing the JPA-off-hot-load and JPA-off-cavity measurements, we can verify the insertion loss measured under real experimental conditions ($\alpha = 0.68\pm 0.05$) between the cavity and the HFET is consistent with the pre-experiment measurement ($\alpha=0.643 \pm 0.003$). Additionally, comparing the JPA-off-hot-load $T_{\rm HFET}/\alpha_{\rm eff}$ ($6.13 \pm 0.20$ K) and the JPA-on-hot-load inferred $T_{\rm HFET}/\alpha_{\rm eff}$ ($6.18 \pm 0.21$ K), we prove that the JPA-Y-factor and the SNRI $\&$ HFET-Y-factor give consistent results. Furthermore, as discussed in the previous section, the ability to perform JPA-on and JPA-off noise calibrations in this calibration campaign provided us with an estimate of the systematic uncertainty on $T_{\rm sys}$ in the axion search analysis for this data-taking run.

\begin{table}[htb!]
    \caption{
   Comparison of different noise calibration measurements at 1280MHz with ADMX.}\label{tab:comparison}
    \begin{ruledtabular}
    \centering
    \begin{tabular}{c c c }
    \textbf{Quantity} & \textbf{Value (K)} & \textbf{Condition}\\\hline
   $T_{\rm HFET}$  & $4.18\pm0.26$ &  JPA-off-cavity \\\hline
    $T_{\rm HFET}/\alpha_{\rm eff}$&  $6.13\pm0.20$ & JPA-off-hot-load \\
     &  $ 6.18 \pm 0.21 $ &JPA-on-hot-load (inferred)  \\
  &  $ 6.72 \pm 0.17 $ & JPA-on-cavity on res. (inferred)\\
     & $ 6.33 \pm 0.21 $& JPA-on-cavity off res. (inferred) \\\hline
    $T_{\rm JPA,eff}$ & $0.141\pm0.014$ & JPA-on-hot-load  \\
    & $0.372\pm0.018$ & JPA-on-cavity on resonance\\\ 
    & $0.372 \pm 0.022$ & JPA-on-cavity off resonance\\
    \end{tabular}
\end{ruledtabular}
\end{table}

\begin{table}[htb!]
    \caption{
   Different physical temperatures for relevant components during the measurements from Sec.~\ref{sec:JPAonhotload} and Sec.~\ref{sec:JPACavityCooldown}.}\label{tab:temperatures}
    \begin{ruledtabular}
    \centering
    \begin{tabular}{c c c c}
    \textbf{Component} & \boldmath{$\rm HL_{\rm Feb}$} & \boldmath{$\rm HL_{\rm June}$}&
    \textbf{Cav. Warm-up}\\\hline
   $T_{\rm cav}$  & 130 mK &  140 mK & 151 mK $\rightarrow 195 $mK \\\hline
  $T_{\rm circ}$  & 95 mK & 100 mK & 119 mK $\rightarrow 180$ mK \\\hline
   $T_{\rm A}$ &  79 mK & 81 mK & 102 mK $\rightarrow 162$ mK  \\\hline
$T_{\rm mxc}$ & 79 mK & 81 mK & 102 mK $\rightarrow  162$ mK \\\hline
   $T_{\rm HL}$ (baseline) & 151 mK & 170 mK & N/A  
    \end{tabular}
\end{ruledtabular}
\end{table}

By performing two JPA-on-hot-load measurements at the same frequency we were able to confirm that the JPA added noise at a given frequency was extremely stable over a long time span, and at two different gains. The gain of the JPA varies during regular data-taking, so it is useful to confirm that the noise performance is not affected by changes in gain on the order of a few dB. Additionally, this data-taking run lasted for nearly a full year so it is important that the noise performance of the JPA did not degrade over time. 

Unexpectedly, the JPA-on-cavity measurements present significantly higher $T_{\rm JPA,eff}$ while compared to the JPA-on-hot-load $T_{\rm JPA,eff}$, which can be confirmed by the SNRI $\&$ HFET-Y-factor $T_{\rm sys}$ measurements in both Fig.~\ref{fig:1280_Tsys_comp_cav_offres} and Fig.~\ref{fig:1280_Tsys_comp_cav_onres}. One possible reason is that the insertion loss between the antenna and the cryo-switch is larger than that between the hot load and the switch, which should be a minor effect for the consistency of $T_{\rm sys}$ using either the direct $T_{\rm JPA,eff}$ fit or SNRI $\&$ $T_{\rm HFET}/\alpha_{\rm eff}$  in Fig.~\ref{fig:1280_Tsys_comp_cav_offres} and Fig.~\ref{fig:1280_Tsys_comp_cav_onres}. Another reason might be that the hot load cannot represent the cavity when it comes to the interaction with the JPA, which is highly possible due to the impedance difference between a hot load (50~$\Omega$) and a cavity (highly-reflective in most frequencies). Lastly, some early observations of the JPA used in these calibration measurements hint that $T_{\rm JPA,eff}$ may have a temperature dependence, with higher physical temperatures of the RF components leading to higher effective JPA noise. As noted in Tab.~\ref{tab:temperatures}, the component temperatures at the high end of the JPA-on-cavity measurement were $\sim 65-85$~mK higher than they were during the hot load measurements, so it is possible that this had an effect on the value of $T_{\rm JPA,eff}$ we measured. Without the addition of JPA-on noise calibration measurements, which are new to this data-taking run, we would not have identified this mysterious, non-negligible increase in the added JPA noise under real experimental circumstances. This will help inform future upgrades for reducing this discrepancy.

At ADMX, to further improve the noise behavior, a lower physical temperature of the milliKelvin space is necessary before reaching the SQL. This could be achieved by using a dilution refrigerator with more cooling power or by implementing experimental design refinements that reduce the overall heat load of the system. A better JPA with lower added noise and higher stable gain can also bring down the system noise. Additionally, a set of circulators with lower insertion loss can be helpful because $\alpha$ will be larger. Circulators with better isolation are also helpful to decrease any standing waves between the cavity and the JPA and potentially reduce the difference in $T_{\rm JPA,eff}$ between switching to the cavity and the hot load.


The consistency in $T_{\rm HFET}/\alpha_{\rm eff}$ between different noise measurements indicates that we can simplify the receiver chain design in future versions of ADMX by removing the cryogenic switch~\cite{Chang_2025}. The simplification can save precious cold and magnetic-free space for other electronic devices as well as further increase the transmissivity between the cavity and JPA and reduce the system noise temperature.

\section{Acknowledgements}
This work was supported by the U.S. Department of Energy through Grants No DE-SC0009800, No. DE-SC0009723, No. DE-SC0010296, No. DE-SC0010280, No. DE-SC0011665, No. DE-FG02-97ER41029, No. DE-FG02-96ER40956, No. DE-AC52-07NA27344, No. DE-AC03-76SF00098, No. DE-SC0022148 and No. DE-SC0017987. This document was prepared by
the ADMX Collaboration using the resources of the
Fermi National Accelerator Laboratory (Fermilab), a
U.S. Department of Energy, Office of Science, Office of
High Energy Physics HEP User Facility. Fermilab is
managed by Fermi Research Alliance, LLC (FRA), acting
under Contract No. DE-AC02-07CH11359. Pacific Northwest National Laboratory is a multi-program national laboratory operated for the U.S. DOE by Battelle Memorial Institute under Contract No. DE-AC05-76RL01830.University of Sheffield acknowledges the Quantum Sensors for the Hidden Sector (QSHS) Extended Support under the grant ST/Y004620/1. Chelsea Bartram acknowledges support from
the Panofsky Fellowship at SLAC. John Clarke acknowledges support from the U.S. Department of Energy, Office of Science, National Quantum Information Science Research Centers. UWA participation is funded by
the ARC Centre of Excellence for Engineered Quantum
Systems, Grant No. CE170100009, Dark Matter Particle
Physics, Grant No. CE200100008, and Forrest Research
Foundation. Additional support was provided by the Heising-Simons Foundation and by the Lawrence Livermore National Laboratory LDRD office. LLNL Release No. LLNL-JRNL-871124. LANL Release No. LA-UR-24-31690.

\bibliographystyle{apsrev4-1}
\raggedright
\bibliography{references.bib}

\end{document}